\documentclass[10pt,conference]{IEEEtran}
\usepackage[utf8]{inputenc}
\usepackage{sourcecodepro}

\usepackage[table,xcdraw,dvipsnames]{xcolor}
\usepackage{nameref}
\usepackage{relsize}
\usepackage{xspace}
\usepackage{lscape}
\usepackage{longtable}
\usepackage{fontawesome5}
\usepackage{listings}
\usepackage{tablefootnote}
\usepackage{environ}
\usepackage{changepage}
\usepackage{textcomp}
\usepackage{pmboxdraw}
\usepackage{tikz}
\usetikzlibrary{shapes.geometric}
\usepackage{multicol}
\usepackage{multirow}
\usepackage[nomessages]{fp}
\usepackage{xparse}
\usepackage{xfp}
\usepackage{siunitx}
\usepackage{subcaption}
\usepackage{hyperref}
\usepackage{xurl}
\usepackage{float}
\usepackage{graphicx}
\usepackage{fancyvrb}
\usepackage{setspace}
\usepackage{tcolorbox}
\tcbuselibrary{listings}
\tcbuselibrary{listingsutf8}
\tcbset{listing engine=listings}
\usepackage{flushend}
\lstdefinelanguage{Rust}{%
  sensitive%
, morecomment=[l]{//}%
, morecomment=[s]{/*}{*/}%
, moredelim=[s][{\itshape\color[rgb]{0,0,0.75}}]{\#[}{]}%
, morestring=[b]{"}%
, alsodigit={}%
, alsoother={}%
, alsoletter={!}%
, morekeywords={int, void, break, continue, else, for, if, in, loop, match, return, while, retag}  
, morekeywords={as, const, let, move, mut, raw, ref, static}  
, morekeywords={dyn, enum, fn, impl, Self, self, struct, trait, type, union, use, wherae}  
, morekeywords={crate, extern, mod, pub, super}  
, morekeywords={unsafe}  
, morekeywords={abstract, alignof, become, box, do, final, macro, offsetof, override, priv, proc, pure, sizeof, typeof, unsized, virtual, yield}  
, morekeywords=[2]{Add, AddAssign, Any, AsciiExt, AsInner, AsInnerMut, AsMut, AsRawFd, AsRawHandle, AsRawSocket, AsRef, Binary, BitAnd, BitAndAssign, Bitor, BitOr, BitOrAssign, BitXor, BitXorAssign, Borrow, BorrowMut, Boxed, BoxPlace, BufRead, BuildHasher, CastInto, CharExt, Clone, CoerceUnsized, CommandExt, Copy, Debug, DecodableFloat, Default, Deref, DerefMut, DirBuilderExt, DirEntryExt, Display, Div, DivAssign, DoubleEndedIterator, DoubleEndedSearcher, Drop, EnvKey, Eq, Error, ExactSizeIterator, ExitStatusExt, Extend, FileExt, FileTypeExt, Float, Fn, FnBox, FnMut, FnOnce, Freeze, From, FromInner, FromIterator, FromRawFd, FromRawHandle, FromRawSocket, FromStr, FullOps, FusedIterator, Generator, Hash, Hasher, Index, IndexMut, InPlace, Int, Into, IntoCow, IntoInner, IntoIterator, IntoRawFd, IntoRawHandle, IntoRawSocket, IsMinusOne, IsZero, Iterator, JoinHandleExt, LargeInt, LowerExp, LowerHex, MetadataExt, Mul, MulAssign, Neg, Not, Octal, OpenOptionsExt, Ord, OsStrExt, OsStringExt, Packet, PartialEq, PartialOrd, Pattern, PermissionsExt, Place, Placer, Pointer, Product, Put, RangeArgument, RawFloat, Read, Rem, RemAssign, Seek, Shl, ShlAssign, Shr, ShrAssign, Sized, SliceConcatExt, SliceExt, SliceIndex, Stats, Step, StrExt, Sub, SubAssign, Sum, Sync, TDynBenchFn, Terminal, Termination, ToOwned, ToSocketAddrs, ToString, Try, TryFrom, TryInto, UnicodeStr, Unsize, UpperExp, UpperHex, WideInt, Write}
, morekeywords=[2]{Send}  
, morekeywords=[3]{bool, char, f32, f64, i8, i16, i32, i64, isize, str, u8, u16, u32, u64, unit, usize, i128, u128}  
, morekeywords=[4]{Err, false, None, Ok, Some, true}  
, morekeywords=[3]{AccessError, Adddf3, AddI128, AddoI128, AddoU128, ADDRESS, ADDRESS64, addrinfo, ADDRINFOA, AddrParseError, Addsf3, AddU128, advice, aiocb, Alignment, AllocErr, AnonPipe, Answer, Arc, Args, ArgsInnerDebug, ArgsOs, Argument, Arguments, ArgumentV1, Ashldi3, Ashlti3, Ashrdi3, Ashrti3, AssertParamIsClone, AssertParamIsCopy, AssertParamIsEq, AssertUnwindSafe, AtomicBool, AtomicPtr, Attr, auxtype, auxv, BackPlace, BacktraceContext, Barrier, BarrierWaitResult, Bencher, BenchMode, BenchSamples, BinaryHeap, BinaryHeapPlace, blkcnt, blkcnt64, blksize, BOOL, boolean, BOOLEAN, BoolTrie, BorrowError, BorrowMutError, Bound, Box, bpf, BTreeMap, BTreeSet, Bucket, BucketState, Buf, BufReader, BufWriter, Builder, BuildHasherDefault, BY, BYTE, Bytes, CannotReallocInPlace, cc, Cell, Chain, CHAR, CharIndices, CharPredicateSearcher, Chars, CharSearcher, CharsError, CharSliceSearcher, CharTryFromError, Child, ChildPipes, ChildStderr, ChildStdin, ChildStdio, ChildStdout, Chunks, ChunksMut, ciovec, clock, clockid, Cloned, cmsgcred, cmsghdr, CodePoint, Color, ColorConfig, Command, CommandEnv, Component, Components, CONDITION, condvar, Condvar, CONSOLE, CONTEXT, Count, Cow, cpu, CRITICAL, CStr, CString, CStringArray, Cursor, Cycle, CycleIter, daddr, DebugList, DebugMap, DebugSet, DebugStruct, DebugTuple, Decimal, Decoded, DecodeUtf16, DecodeUtf16Error, DecodeUtf8, DefaultEnvKey, DefaultHasher, dev, device, Difference, Digit32, DIR, DirBuilder, dircookie, dirent, dirent64, DirEntry, Discriminant, DISPATCHER, Display, Divdf3, Divdi3, Divmoddi4, Divmodsi4, Divsf3, Divsi3, Divti3, dl, Dl, Dlmalloc, Dns, DnsAnswer, DnsQuery, dqblk, Drain, DrainFilter, Dtor, Duration, DwarfReader, DWORD, DWORDLONG, DynamicLibrary, Edge, EHAction, EHContext, Elf32, Elf64, Empty, EmptyBucket, EncodeUtf16, EncodeWide, Entry, EntryPlace, Enumerate, Env, epoll, errno, Error, ErrorKind, EscapeDebug, EscapeDefault, EscapeUnicode, event, Event, eventrwflags, eventtype, ExactChunks, ExactChunksMut, EXCEPTION, Excess, ExchangeHeapSingleton, exit, exitcode, ExitStatus, Failure, fd, fdflags, fdsflags, fdstat, ff, fflags, File, FILE, FileAttr, filedelta, FileDesc, FilePermissions, filesize, filestat, FILETIME, filetype, FileType, Filter, FilterMap, Fixdfdi, Fixdfsi, Fixdfti, Fixsfdi, Fixsfsi, Fixsfti, Fixunsdfdi, Fixunsdfsi, Fixunsdfti, Fixunssfdi, Fixunssfsi, Fixunssfti, Flag, FlatMap, Floatdidf, FLOATING, Floatsidf, Floatsisf, Floattidf, Floattisf, Floatundidf, Floatunsidf, Floatunsisf, Floatuntidf, Floatuntisf, flock, ForceResult, FormatSpec, Formatted, Formatter, Fp, FpCategory, fpos, fpos64, fpreg, fpregset, FPUControlWord, Frame, FromBytesWithNulError, FromUtf16Error, FromUtf8Error, FrontPlace, fsblkcnt, fsfilcnt, fsflags, fsid, fstore, fsword, FullBucket, FullBucketMut, FullDecoded, Fuse, GapThenFull, GeneratorState, gid, glob, glob64, GlobalDlmalloc, greg, group, GROUP, Guard, GUID, Handle, HANDLE, Handler, HashMap, HashSet, Heap, HINSTANCE, HMODULE, hostent, HRESULT, id, idtype, if, ifaddrs, IMAGEHLP, Immut, in, in6, Incoming, Infallible, Initializer, ino, ino64, inode, input, InsertResult, Inspect, Instant, int16, int32, int64, int8, integer, IntermediateBox, Internal, Intersection, intmax, IntoInnerError, IntoIter, IntoStringError, intptr, InvalidSequence, iovec, ip, IpAddr, ipc, Ipv4Addr, ipv6, Ipv6Addr, Ipv6MulticastScope, Iter, IterMut, itimerspec, itimerval, jail, JoinHandle, JoinPathsError, KDHELP64, kevent, kevent64, key, Key, Keys, KV, l4, LARGE, lastlog, launchpad, Layout, Lazy, lconv, Leaf, LeafOrInternal, Lines, LinesAny, LineWriter, linger, linkcount, LinkedList, load, locale, LocalKey, LocalKeyState, Location, lock, LockResult, loff, LONG, lookup, lookupflags, LookupHost, LPBOOL, LPBY, LPBYTE, LPCSTR, LPCVOID, LPCWSTR, LPDWORD, LPFILETIME, LPHANDLE, LPOVERLAPPED, LPPROCESS, LPPROGRESS, LPSECURITY, LPSTARTUPINFO, LPSTR, LPVOID, LPWCH, LPWIN32, LPWSADATA, LPWSAPROTOCOL, LPWSTR, Lshrdi3, Lshrti3, lwpid, M128A, mach, major, Map, mcontext, Metadata, Metric, MetricMap, mflags, minor, mmsghdr, Moddi3, mode, Modsi3, Modti3, MonitorMsg, MOUNT, mprot, mq, mqd, msflags, msghdr, msginfo, msglen, msgqnum, msqid, Muldf3, Mulodi4, Mulosi4, Muloti4, Mulsf3, Multi3, Mut, Mutex, MutexGuard, MyCollection, n16, NamePadding, NativeLibBoilerplate, nfds, nl, nlink, NodeRef, NoneError, NonNull, NonZero, nthreads, NulError, OccupiedEntry, off, off64, oflags, Once, OnceState, OpenOptions, Option, Options, OptRes, Ordering, OsStr, OsString, Output, OVERLAPPED, Owned, Packet, PanicInfo, Param, ParseBoolError, ParseCharError, ParseError, ParseFloatError, ParseIntError, ParseResult, Part, passwd, Path, PathBuf, PCONDITION, PCONSOLE, Peekable, PeekMut, Permissions, PhantomData, pid, Pipes, PlaceBack, PlaceFront, PLARGE, PoisonError, pollfd, PopResult, port, Position, Powidf2, Powisf2, Prefix, PrefixComponent, PrintFormat, proc, Process, PROCESS, processentry, protoent, PSRWLOCK, pthread, ptr, ptrdiff, PVECTORED, Queue, radvisory, RandomState, Range, RangeFrom, RangeFull, RangeInclusive, RangeMut, RangeTo, RangeToInclusive, RawBucket, RawFd, RawHandle, RawPthread, RawSocket, RawTable, RawVec, Rc, ReadDir, Receiver, recv, RecvError, RecvTimeoutError, ReentrantMutex, ReentrantMutexGuard, Ref, RefCell, RefMut, REPARSE, Repeat, Result, Rev, Reverse, riflags, rights, rlim, rlim64, rlimit, rlimit64, roflags, Root, RSplit, RSplitMut, RSplitN, RSplitNMut, RUNTIME, rusage, RwLock, RWLock, RwLockReadGuard, RwLockWriteGuard, sa, SafeHash, Scan, sched, scope, sdflags, SearchResult, SearchStep, SECURITY, SeekFrom, segment, Select, SelectionResult, sem, sembuf, send, Sender, SendError, servent, sf, Shared, shmatt, shmid, ShortReader, ShouldPanic, Shutdown, siflags, sigaction, SigAction, sigevent, sighandler, siginfo, Sign, signal, signalfd, SignalToken, sigset, sigval, Sink, SipHasher, SipHasher13, SipHasher24, size, SIZE, Skip, SkipWhile, Slice, SmallBoolTrie, sockaddr, SOCKADDR, sockcred, Socket, SOCKET, SocketAddr, SocketAddrV4, SocketAddrV6, socklen, speed, Splice, Split, SplitMut, SplitN, SplitNMut, SplitPaths, SplitWhitespace, spwd, SRWLOCK, ssize, stack, STACKFRAME64, StartResult, STARTUPINFO, stat, Stat, stat64, statfs, statfs64, StaticKey, statvfs, StatVfs, statvfs64, Stderr, StderrLock, StderrTerminal, Stdin, StdinLock, Stdio, StdioPipes, Stdout, StdoutLock, StdoutTerminal, StepBy, String, StripPrefixError, StrSearcher, subclockflags, Subdf3, SubI128, SuboI128, SuboU128, subrwflags, subscription, Subsf3, SubU128, Summary, suseconds, SYMBOL, SYMBOLIC, SymmetricDifference, SyncSender, sysinfo, System, SystemTime, SystemTimeError, Take, TakeWhile, tcb, tcflag, TcpListener, TcpStream, TempDir, TermInfo, TerminfoTerminal, termios, termios2, TestDesc, TestDescAndFn, TestEvent, TestFn, TestName, TestOpts, TestResult, Thread, threadattr, threadentry, ThreadId, tid, time, time64, timespec, TimeSpec, timestamp, timeval, timeval32, timezone, tm, tms, ToLowercase, ToUppercase, TraitObject, TryFromIntError, TryFromSliceError, TryIter, TryLockError, TryLockResult, TryRecvError, TrySendError, TypeId, U64x2, ucontext, ucred, Udivdi3, Udivmoddi4, Udivmodsi4, Udivmodti4, Udivsi3, Udivti3, UdpSocket, uid, UINT, uint16, uint32, uint64, uint8, uintmax, uintptr, ulflags, ULONG, ULONGLONG, Umoddi3, Umodsi3, Umodti3, UnicodeVersion, Union, Unique, UnixDatagram, UnixListener, UnixStream, Unpacked, UnsafeCell, UNWIND, UpgradeResult, useconds, user, userdata, USHORT, Utf16Encoder, Utf8Error, Utf8Lossy, Utf8LossyChunk, Utf8LossyChunksIter, utimbuf, utmp, utmpx, utsname, uuid, VacantEntry, Values, ValuesMut, VarError, Variables, Vars, VarsOs, Vec, VecDeque, vm, Void, WaitTimeoutResult, WaitToken, wchar, WCHAR, Weak, whence, WIN32, WinConsole, Windows, WindowsEnvKey, winsize, WORD, Wrapping, wrlen, WSADATA, WSAPROTOCOL, WSAPROTOCOLCHAIN, Wtf8, Wtf8Buf, Wtf8CodePoints, xsw, xucred, Zip, zx}
, morekeywords=[5]{assert!, assert_eq!, assert_ne!, cfg!, column!, compile_error!, concat!, concat_idents!, debug_assert!, debug_assert_eq!, debug_assert_ne!, env!, eprint!, eprintln!, file!, format!, format_args!, include!, include_bytes!, include_str!, line!, module_path!, option_env!, panic!, print!, println!, select!, stringify!, thread_local!, try!, unimplemented!, unreachable!, vec!, write!, writeln!}  
}%

\lstdefinestyle{colouredRust}%
{ basicstyle=\ttfamily\small%
, identifierstyle=\color[rgb]{0, 0.5, 0}%
, commentstyle=\color[gray]{0.4}%
, stringstyle=\color[rgb]{0.63, 0.08, 0.08}%
, numberstyle=\ttfamily\scriptsize\color{black} 
, keywordstyle=\bfseries\color[rgb]{0.05, 0.05, 0.98}
, keywordstyle=[2]\color[rgb]{0.05, 0.05, 0.98}
, keywordstyle=[3]\color[rgb]{0, 0.5, 0}
, keywordstyle=[4]\color[rgb]{0, 0.5, 0}
, keywordstyle=[5]\color[rgb]{0, 0, 0.75}
, keepspaces=truef%
, showspaces=false%
, showtabs=false%
, showstringspaces=true%
, numbers=left%
}%

\lstdefinestyle{boxed}{
  style=colouredRust%
, numbers=left%
, firstnumber=auto%
, numberblanklines=true%
, frame=trbL%
, numberstyle=%
, frame=leftline%
, numbersep=7pt%
, framesep=5pt%
, framerule=10pt%
, xleftmargin=15pt%
, backgroundcolor=%
, rulecolor=\color[gray]{1}%
}
\input{highlighter.tex}
\hypersetup{breaklinks=true}

\newcommand*\circled[1]{\tikz[baseline=(char.base)]{
\node[shape=circle,draw,inner sep=.8pt] (char) {#1};}}

\newcommand*\rectangled[1]{\tikz[baseline=(char.base)]{
\node[shape=rectangle,draw,inner sep=1.6pt] (char) {#1};}}

\definecolor{hlcolorred}{RGB}{255, 218, 214}
\definecolor{hlcolorgreen}{RGB}{216, 232, 220}

\newcounter{snippetcounter}
\lstdefinestyle{rustcodestyle}{
  language=Rust, 
  style=colouredRust, 
  escapeinside={|}{|},
  moredelim=**[is][{\btHL[fill=hlcolorred]}]{<<^}{^>>},
  moredelim=**[is][{\btHL[fill=hlcolorgreen]}]{<<@}{@>>},  
}

\let\origthelstnumber\thelstnumber
\makeatletter
\newcommand*\Suppressnumber{%
  \lst@AddToHook{OnNewLine}{%
    \let\thelstnumber\relax%
     \advance\c@lstnumber-\@ne\relax%
    }%
}

\newcommand*\Reactivatenumber{%
  \lst@AddToHook{OnNewLine}{%
   \let\thelstnumber\origthelstnumber%
   \advance\c@lstnumber\@ne\relax}%
}

\lstnewenvironment{rustfigure}{
  \lstset{
    style=rustcodestyle,
    basicstyle=\ttfamily\scriptsize,
    xleftmargin=0.75cm,
  }
}{}
\lstnewenvironment{rustfigurenoln}{
  \lstset{
    style=rustcodestyle, 
    numbers=none, 
    basicstyle=\ttfamily\scriptsize,
    xleftmargin=10pt,
  }
}{}

\newcommand{\code}[1]{\lstinline[style=rustcodestyle]|#1|}

\newcommand{\unsafe}{\code{unsafe}\xspace}

\newcounter{bugcounter}
\newcommand{\refbug}[1]{\#\getrefnumber{#1}}

\newcommand{\toolname}{MiriLLI\xspace}

\newcommand{\citecpp}[1]{C++23\ \S#1}
\newcommand{\citecc}[1]{C23\ \S#1}

\newcommand{\m}[1]{\mathsf{#1}}
\newcommand{\tbReserved}{$\m{Reserved}$\xspace}
\newcommand{\tbActive}{$\m{Active}$\xspace}
\newcommand{\tbFrozen}{$\m{Frozen}$\xspace}
\newcommand{\tbDisabled}{$\m{Disabled}$\xspace}
\refstepcounter{bugcounter}\label{quickjs:regex:3}
\refstepcounter{bugcounter}\label{everrs:1}
\refstepcounter{bugcounter}\label{quickjs:regex:1}
\refstepcounter{bugcounter}\label{secp256k1:1}
\refstepcounter{bugcounter}\label{spritz:cipher:1}
\refstepcounter{bugcounter}\label{special-fun:1}
\refstepcounter{bugcounter}\label{dec-number-sys:2}
\refstepcounter{bugcounter}\label{lcms2:2}
\refstepcounter{bugcounter}\label{bchlib:1}
\refstepcounter{bugcounter}\label{libcmark-sys:1}
\refstepcounter{bugcounter}\label{tinyspline-sys:1}
\refstepcounter{bugcounter}\label{x42ltc-sys:1}
\refstepcounter{bugcounter}\label{minimp3:ex-sys:1}
\refstepcounter{bugcounter}\label{fluidlite:1}
\refstepcounter{bugcounter}\label{ytnef:1}
\refstepcounter{bugcounter}\label{zmq:1}
\refstepcounter{bugcounter}\label{libhydrogen:1}
\refstepcounter{bugcounter}\label{crypto:pimitives:1}
\refstepcounter{bugcounter}\label{dec:2}
\refstepcounter{bugcounter}\label{dec-number-sys:3}
\refstepcounter{bugcounter}\label{quickjs:regex:2}
\refstepcounter{bugcounter}\label{mseed:1}
\refstepcounter{bugcounter}\label{bad64:1}
\refstepcounter{bugcounter}\label{rsofa:1}
\refstepcounter{bugcounter}\label{tectonic:engine:bibtex:1}
\refstepcounter{bugcounter}\label{sgp4-rs:1}
\refstepcounter{bugcounter}\label{lsmlite-rs:1}
\refstepcounter{bugcounter}\label{ms5837:1}
\refstepcounter{bugcounter}\label{dec-number-sys:1}
\refstepcounter{bugcounter}\label{foreign-types:1}
\refstepcounter{bugcounter}\label{bzip2:1}
\refstepcounter{bugcounter}\label{littlefs2:1}
\refstepcounter{bugcounter}\label{littlefs2:2}
\refstepcounter{bugcounter}\label{spng:1}
\refstepcounter{bugcounter}\label{flate2:1}
\refstepcounter{bugcounter}\label{librsync:1}
\refstepcounter{bugcounter}\label{dec:1}
\refstepcounter{bugcounter}\label{blitsort-sys:1}
\refstepcounter{bugcounter}\label{lcms2:1}
\refstepcounter{bugcounter}\label{lcms2:3}
\refstepcounter{bugcounter}\label{klu-rs:1}
\refstepcounter{bugcounter}\label{minimap2-sys:1}
\refstepcounter{bugcounter}\label{clickhouse-driver-lz4:1}
\refstepcounter{bugcounter}\label{speex-safe:1}
\refstepcounter{bugcounter}\label{dec:3}
\refstepcounter{bugcounter}\label{xxhrs:1}

\ExplSyntaxOn
\ior_new:N \l__lrnv_temp_ior
\str_new:N \l__lrnv_tmp_str
\seq_new:N \l__lrnv_key_seq
\seq_new:N \l__lrnv_value_seq
\prop_new:N \l__lrnv_array_prop
\cs_new_protected:Npn \__lrnv_csv_to_prop:n #1
  {
    \__lrnv_input_csv:n {#1}
    \seq_mapthread_function:NNN \l__lrnv_key_seq \l__lrnv_value_seq
      \__lrnv_set_prop:nn
  }
\cs_new_protected:Npn \__lrnv_set_prop:nn #1 #2
  { \prop_put:Nnn \l__lrnv_array_prop {#1} {#2} }
\cs_new_protected:Npn \__lrnv_input_csv:n #1
  {
    \ior_open:Nn \l__lrnv_temp_ior {#1}
    \__lrnv_read_csv_row_to_seq:NN \l__lrnv_temp_ior \l__lrnv_key_seq
    \__lrnv_read_csv_row_to_seq:NN \l__lrnv_temp_ior \l__lrnv_value_seq
    \ior_close:N \l__lrnv_temp_ior
  }
\cs_new_protected:Npn \__lrnv_read_csv_row_to_seq:NN #1 #2
  {
    \ior_str_get:NN #1 \l__lrnv_tmp_str
    \seq_set_from_clist:NV #2 \l__lrnv_tmp_str
  }
\cs_generate_variant:Nn \seq_set_from_clist:Nn { NV }
\msg_new:nnn { lrnv } { file-not-found }
  { File~`#1'~not~found. }
\NewDocumentCommand \ReadCSVArray { m }
  {
    \file_if_exist:nTF {#1}
      { \__lrnv_csv_to_prop:n {#1} }
      { \msg_error:nnn { lrnv } { file-not-found } {#1} }
  }
\NewExpandableDocumentCommand \ArrayItem { m }
  { \prop_item:Nn \l__lrnv_array_prop { "#1" } }
\ExplSyntaxOff

\renewcommand{\footnotemark}{\mbox{}}
\def\BibTeX{{\rm B\kern-.05em{\sc i\kern-.025em b}\kern-.08em
    T\kern-.1667em\lower.7ex\hbox{E}\kern-.125emX}}

\begin{document}
\title{A Study of Undefined Behavior Across Foreign Function Boundaries in Rust Libraries} 
\author{\IEEEauthorblockN{Ian McCormack}
\IEEEauthorblockA{Carnegie Mellon University\\
Pittsburgh, PA, USA \\
icmccorm@cs.cmu.edu}
\and
\IEEEauthorblockN{Joshua Sunshine}
\IEEEauthorblockA{Carnegie Mellon University\\
Pittsburgh, PA, USA \\
sunshine@cs.cmu.edu}
\and
\IEEEauthorblockN{Jonathan Aldrich}
\IEEEauthorblockA{Carnegie Mellon University \\
Pittsburgh, PA, USA \\
jonathan.aldrich@cs.cmu.edu}
}

\maketitle
\pagestyle{plain}

\begin{abstract}
Developers rely on the static safety guarantees of the Rust programming language to write secure and performant applications. However, Rust is frequently used to interoperate with other languages which allow design patterns that conflict with Rust's evolving aliasing models. Miri is currently the only dynamic analysis tool that can validate applications against these models, but it does not support finding bugs in foreign functions, indicating that there may be a critical correctness gap across the Rust ecosystem. We conducted a large-scale evaluation of Rust libraries that call foreign functions to determine whether Miri's dynamic analyses remain useful in this context. We used Miri and an LLVM interpreter to jointly execute applications that call foreign functions, where we found 46 instances of undefined or undesired behavior in 37 libraries. Three bugs were found in libraries that had more than 10,000 daily downloads on average during our observation period, and one was found in a library maintained by the Rust Project. Many of these bugs were violations of Rust's aliasing models, but the latest Tree Borrows model was significantly more permissive than the earlier Stacked Borrows model. The Rust community must invest in new, production-ready tooling for multi-language applications to ensure that developers can detect these errors.
\end{abstract}


\section{Introduction}
\label{section:introduction}
The Rust programming language has become increasingly popular due to its static safety guarantees, which provide security benefits comparable to garbage collection without additional run-time overhead~\cite{devsurvey, pereira17}.
However, Rust is also frequently used in interoperation with languages that do not provide similar assurances. To call foreign functions, developers must use a subset of unsafe features to bypass Rust's restrictions. If these features are used incorrectly, they can break Rust's aliasing rules. The Rust compiler relies on these rules to optimize code. If they are broken, optimizations may be applied incorrectly, which can introduce security vulnerabilities. 

Miri~\cite{miri} is a widely-used Rust interpreter that uses dynamic analysis to detect violations of Rust's aliasing model. A limitation of Miri is that it cannot detect these errors across foreign function boundaries. However, one of the most common reasons for using Rust's unsafe features is to call foreign functions~\cite{evans20,astrauskas20,fulton21,holtervennhoff23,mccormack24}. We seek to determine whether the differences between Rust and other languages are leading to errors in practice.
\begin{itemize}
\item\textbf{RQ1}: What types of errors occur in Rust libraries that call foreign functions?
\end{itemize}
The Rust community has proposed two aliasing models: Stacked Borrows~\cite{stackedborrows} and Tree Borrows~\cite{treeborrows}. The goal of these models is to "strike a balance"~\cite{stackedborrows} between performance and usability by providing a set of rules that developers must follow to ensure that compile-time optimizations are applied correctly~\cite{jung21}. Since Stacked Borrows and Tree Borrows both provide rules of this kind, we ask a second research question: 
\begin{itemize}
\item\textbf{RQ2}: Which of Rust's aliasing models permits more real-world programs with foreign function calls?
\end{itemize}
To answer these questions, we created \toolname: a tool which combines Miri with an LLVM interpreter to jointly execute programs and detect undefined behavior across foreign function boundaries.
We used \toolname to conduct a large-scale study of 9,130 test cases from 957 Rust libraries that call foreign functions. We identified 46 unique instances of undefined or undesirable behavior from 37 libraries. Of the 90 test cases that violated Stacked Borrows, 66\% (59) did \textit{not} violate Tree Borrows.

Our results indicate that Rust's restrictions on aliasing, mutability, and initialization make it easy to inadvertently introduce undefined behavior when calling foreign functions. Developers can take immediate steps to avoid these errors by auditing their use of certain types at foreign callsites. However, the Rust Project must invest in new, production-ready tooling to ensure that these errors can be easily detected. 

\paragraph*{Overview}
In Section~\ref{section:background}, we compare Rust's semantics with C and C++, and we describe the resources and best practices that Rust developers use to interoperate with these languages.
In Section~\ref{section:methodology}, we document our methodology for sampling and evaluating test cases from Rust libraries that call foreign functions, and we describe the challenges that we encountered when implementing \toolname.
In Section~\ref{section:results}, we describe each type of bug that we found.
We discuss the implications of our findings in Section~\ref{section:discussion}.
We review prior work on Rust interoperation in Section~\ref{section:related}, we discuss threats to validity in Section~\ref{section:threats}, and we conclude in Section~\ref{section:conclusion}. Our dataset, the Appendix, and the source for \toolname are available in our replication package\footnote{\url{https://doi.org/10.5281/zenodo.12727039}}.

\section{Background}
\label{section:background}
Rust's safety restrictions begin at the level of a value, which is valid for a particular scope. A value's type implements \textit{traits} that define how it behaves. For example, all types have \textit{move semantics} by default, meaning that each value has a unique owner.
However, values with the \code{Copy} trait have no owner and can be freely duplicated.
Ownership can be transferred through assignment or \textit{borrowed} by creating a reference.
References are either mutable, taking the form \code{\&mut T}, or immutable, with the form \code{\&T}.
Mutable references have move semantics, but immutable references can be copied.
Each reference has a \textit{lifetime}, which is the portion of the program over which it will be used.
Rust's borrow checker statically enforces that the lifetime of a reference must not exceed the scope or lifetime of the value that it borrows.
A value can have a single mutable reference or many immutable references that are active within a given context~\cite{crichton20}, but not both at the same time. 

These restrictions prevent safety issues, but they can also make it impossible to implement certain design patterns. When developers need to bypass Rust's restrictions, they can use the \unsafe keyword to enable a set of additional features that are not restricted by the borrow checker. 
These features include dereferencing raw pointers, accessing the fields of union types, modifying static mutable state, implementing unsafe traits, and calling unsafe functions---including those written in other languages~\cite{rustbook}.
The Rust compiler can assume that programs that use unsafe code will follow the rules of its aliasing model.
Programs that violate these rules have \textit{undefined behavior}, so they may be optimized incorrectly. This can lead to differences in behavior that can cause security vulnerabilities~\cite{cve_2023_30624} in practice.

\subsection{Rust's Aliasing Model}
\label{background:borrows}
Rust developers can avoid these issues by ensuring that their programs adhere to the rules of Rust's aliasing model. The Tree Borrows~\cite{treeborrows} model provides the latest definition of these rules. Under this model, each pointer has \textit{provenance} that determines its permission to read or write to a location~\cite{jung24}. Each location is associated with a tree that tracks all valid permissions to its contents. When a location is borrowed, a branch is created in the tree that holds a new permission to access the location. When a pointer is used, its permission is checked against the tree to ensure that it is valid and that it permits the kind of access that takes place. Each type of access will have a different effect on the tree depending on the position of each permission in the tree relative to the permission used for the access. With respect to a given permission, an access is considered a \textit{child access} if it requires that permission or any of its descendants; otherwise, it is a \textit{foreign access}. Accessing a location using an invalid permission is undefined behavior. 

There are three categories of aliasing violations under Tree Borrows that we observed in our investigation. Figure~\ref{figures:borrows} provides minimal examples of each one. We refer to the first category as an \textit{expired permission} error. Under Tree Borrows, writing to a memory location may cause other permissions to that location to expire (or transition to \tbDisabled, in terms of the model). Using one of these expired permissions is undefined behavior. We provide a minimal example of this in Figure~\ref{expired:firstmut}, where the variable \code{x} is mutably borrowed twice; once by \code{y} and again by \code{z}. However, we also store a copy of \code{y} as a raw pointer (\code{*mut}). Rust's borrow checker does not place aliasing restrictions on raw pointers, so it assumes that \code{z} has exclusive, mutable access from when it is created on line~\ref{expired:secondmut} until it is used on line~\ref{expired:ub}. Writing through \code{y} on line~\ref{expired:write} breaks this assumption, so writing through \code{z} is now undefined behavior. In terms of the model, the permissions for \code{y} and \code{z} are children of the permission for \code{x}. Both receive a \tbReserved permission, which has the capability to become a unique \tbActive permission for writing to \code{x}. When a \tbReserved permission is used for a write access, it becomes an \tbActive permission, and all foreign \tbReserved permissions become \tbDisabled, canceling their reservations. Since \code{z} is adjacent to \code{y}, the write on line~\ref{expired:write} is a foreign access. 
\begin{figure}
\scriptsize
\centering
\caption{Minimal examples of the categories of Stacked and Tree Borrows violations that we observed in our evaluation. Statements that trigger undefined behavior are highlighted in red.}
\label{figures:borrows}
\begin{subfigure}[t]{\columnwidth}
\caption{A minimal example of an \textit{expired permission} error. The table on the righthand side shows the tree of permissions for the local variable \code{x} before and after writing with \code{y} on line~\ref{expired:write}.}
\label{figures:borrows:expired}
\begin{minipage}[t]{0.49\columnwidth}
\vspace{0pt}
\begin{rustfigure}
let mut x = 0;|\label{expired:decl}|
let y = &mut x;|\label{expired:firstmut}|
let y = y as *mut _;|\label{expired:saveraw}|
let z = &mut x;|\label{expired:secondmut}|
unsafe { *y = 1; }|\label{expired:write}|
<<^*z = 0;^>>|\label{expired:ub}|
\end{rustfigure}\end{minipage}\begin{minipage}[t]{0.5\columnwidth}
\vspace{0pt}

{
\renewcommand{\baselinestretch}{1.2}\selectfont
$$
\begin{array}{|ccc|l}
  \cline{1-3} \multicolumn{3}{|c|}{\m{Active}} & \verb|└┬|\text{\code{x}}\\ \cline{1-3}
  \m{Reserved} &\rightarrow& \m{Active}        & \verb| ├─|{\text{\code{y}}} \\ \cline{1-3}
  \m{Reserved} &\rightarrow& \m{Disabled}      & \verb| └─|\text{\code{z}} \\ \cline{1-3}
  \end{array}
$$
}
\end{minipage}

\end{subfigure}\hfill\begin{subfigure}[t]{\columnwidth}
\caption{A minimal example of an \textit{insufficient permission} error.}
\label{figures:borrows:insufficient}
\begin{rustfigure}
let x = 0;
ub(&x);|\label{figures:borrows:insufficient:cast}|
fn ub(x: *const u8) {
 let x = x as *mut _;
 unsafe { <<^*x = 1^>> };|\label{figures:borrows:insufficient:write}|
}
\end{rustfigure}
\end{subfigure}
\begin{subfigure}[t]{\columnwidth}
\caption{A minimal example of a \textit{protected permission} error.}
\label{figures:borrows:protection}
\begin{rustfigure}
unsafe fn free(x: &mut u8, layout: Layout) {
  dealloc(x as *mut _, layout);
}
\end{rustfigure}
\end{subfigure}
\begin{subfigure}[t]{\columnwidth}
\caption{A minimal example of an \textit{access out-of-bounds} error under Stacked Borrows that is accepted under Tree Borrows.}
\label{figures:borrows:range}
\begin{rustfigure}
let x: &mut i32 = &mut (0, 0).0;
let x = (x as *mut i32).offset(1);
unsafe { <<^*x = 1^>> };|\label{range:oob}|
\end{rustfigure}
\end{subfigure}
\end{figure}

The second category of undefined behavior is an \textit{insufficient permission} error. This type of error occurs when a reference or pointer with a read-only permission is used to write to memory.
This occurs in Figure~\ref{figures:borrows:insufficient} on line~\ref{figures:borrows:insufficient:write}, where the function \code{ub} writes through a pointer that was cast from an immutable reference (\code{\&x}) on line~\ref{figures:borrows:insufficient:cast}. Under Tree Borrows, this reference is given a \tbFrozen permission, which is read-only. Casting a reference into a raw pointer does not change its permission, and neither does casting a pointer between \code{const} and \code{mut}.

The third type of undefined behavior that we observed is a \textit{protected permission} error. We provide an example in Figure~\ref{figures:borrows:protection}, where the function \code{free} receives a mutable reference. This reference encodes the promise that ownership will eventually be returned to the caller. However, this promise is broken when the reference is cast into a raw pointer and used to free the underlying allocation. In terms of Tree Borrows, when a reference-type value is passed as an argument to a function, it becomes \textit{protected}, making it undefined behavior for it to transition to \tbDisabled---even if it is never accessed after this point. When a function returns, the permissions of its arguments are no longer protected.

\subsubsection{Differences from Stacked Borrows}
Tree Borrows replaces the Stacked Borrows~\cite{stackedborrows} model, which uses a stack to track the active permissions to each location. Under Stacked borrows, a reference cannot be offset and used to access locations outside of the range it originally borrowed. Figure~\ref{figures:borrows:range} shows an example of this \textit{access out-of-bounds} error; the dereference on line~\ref{range:oob} is invalid, since \code{x} only borrowed the first element of the tuple. This pattern is allowed under Tree Borrows; a valid permission can be used to access any location within the bounds of its allocation. Tree Borrows also delays asserting that the permission of a reference is valid until it is used for the first time, whereas Stacked Borrows asserts that a reference is valid immediately when it is created. However, Tree Borrows is less permissive than Stacked Borrows for certain aliasing patterns due to the semantics of its \tbReserved permission, justifying our second research question.

\subsection{Rust vs. C and C++}
\label{background:rustvs}
Rust shares several other categories of undefined behavior with current C~\cite{c23} and C++~\cite{cpp23} standards, such as accessing memory that has been freed, accessing beyond the bounds of an allocation, creating a data race, reading uninitialized memory, and accessing a value at an unaligned address~\cite{rustref}. However, C and C++ have different rules related to initialization, aliasing, and mutability that may conflict with Rust's expectations.

\subsubsection{Initialization}
In both C and C++, accessing an uninitialized object produces an ``indeterminate value'', which is undefined behavior.\footnote{\citecc{6.2.4.6-7};\ \citecpp{6.7.4.1}} However, there are a few exceptions. Static and thread-local variables are initialized immediately at the beginning of their lifetime, and both the \code{unsigned char} and \code{byte} types are allowed to store indeterminate values. It is also safe to create a reference to an uninitialized value.

Rust's borrow checker prevents variables from being used until they are initialized, and no primitive types tolerate indeterminate values in safe contexts. However, Rust's standard library provides \code{MaybeUninit<T>}, which represents an instance of \code{T} that may or may not be initialized. Once an instance has been fully initialized, the unsafe function \code{MaybeUninit<T>::assume_init} will unwrap the outer struct to produce the value inside. It is undefined behavior to call \code{assume_init} if \code{T} is not fully initialized. 

\subsubsection{Aliasing \& Mutability} Neither C nor C++ statically restricts aliasing, but the standards for each language allow implementations to make type-directed assumptions about aliasing. In both languages, it is considered undefined behavior for a pointer to one type to refer to a value of another type if the two types differ.\footnote{\citecc{6.5.7};\ \citecpp{7.2.1.11}} However, differences do not include qualifiers, so two parameters of types \code{int *} and \code{const int *} can alias.

In Rust, both variables and references have distinct, connected capabilities for mutation. Variables are immutable by default, only mutable variables can be mutably borrowed, and only mutable references can be used for mutation. In C and C++, variables are \textit{mutable} by default, and the capability to mutate memory is determined by the mutability of the object being pointed to, regardless of the pointer's type.\footnote{\citecc{6.7.4.1.7};\ \citecpp{9.2.9.2.3}} If a variable was declared as mutable, then a \code{const} pointer can be cast into a mutable pointer and used to write to it.

\subsubsection{Provenance} 
Both C and C++ allow pointers to be converted to and from integers with a size equal to the word size of the current architecture.\footnote{\citecc{6.3.2.3.5-6};\ \citecpp{6.7.1.10.4-5}} Implementations may track the provenance, or origin of pointers to inform compile-time optimizations~\cite{provenance}. However, neither of the current standards for C or C++ defines if or how provenance should be preserved across these conversions, though several models have been proposed~\cite{memarian19}. Rust's pointers have provenance~\cite{jung24}, but the specifics of Rust's provenance model are undecided.

Miri has multiple methods for handling provenance across pointer-to-integer conversion. By default, when a pointer is converted to an integer, its tag is added to a set of tags that have been ``exposed'' by pointer-to-integer conversion. When integers are converted back into pointers, their allocation identifier is reconstructed from their address, but they receive a ``wildcard'' provenance value~\cite{jung22}. Under Stacked Borrows, when a memory access occurs using a wildcard tag, Miri eagerly interprets it as being equivalent to any tag in the stack that permits the access. The current implementation of Tree Borrows does not support this behavior yet; all accesses with wildcard tags are allowed and do not affect the state of the tree. Alternatively, developers can enable \textit{strict provenance}, which treats integer-to-pointer conversion as an error.

\subsection{Interoperation}
\label{background:interop}
Developers can access foreign functions and static variables using Rust's foreign function interface. Declarations, or ``bindings,'' for each of these objects are written in \code{extern} blocks, which take an optional qualifier to indicate the Application Binary Interface (ABI) used by the foreign library. Developers can also use the \code{extern} keyword to change the ABI of a function defined in Rust so that it can be called from other languages. Only a subset of Rust's types are guaranteed to be compatible with foreign ABIs. Both structs and enums must be annotated with the \code{#[repr(C)]} attribute to ensure that their layout is compatible with C. Bindings are not validated against their definitions, but tools such as bindgen~\cite{bindgen} and CXX~\cite{cxx} can automatically generate bindings from headers.

Several dynamic analysis tools can find undefined behavior in multi-language Rust applications. Valgrind~\cite{seward05} is capable of detecting spatial and temporal memory errors, as well as incompatibilities with size and alignment. Many of the LLVM Project's sanitizers are also compatible with Rust~\cite{rust_sanitizers}. Each of these tools is available as a plugin for Cargo, Rust's build tool. Miri can call foreign functions from natively compiled shared libraries using libffi~\cite{libffi}, but it does not support passing certain argument types to native functions, and it cannot find undefined behavior that is triggered in foreign code.

\section{Methodology}
\label{section:methodology} 
We seek to determine whether the differences between Rust and other languages lead to undefined behavior in practice. 
First, we analyzed all libraries published on \href{https://crates.io/}{crates.io}---Rust's central package repository---to find the subset with test cases that produce LLVM bitcode for C or C++ libraries during their build process. We describe this stage in Section~\ref{method:sampling}.
Then, we created \toolname, which can run these test cases and detect undefined behavior. 
\toolname extends Miri to interoperate with LLI~\cite{lli}, an LLVM interpreter. Both Miri and LLI are included within the Rust toolchain; LLI is part of the LLVM backend. \toolname uses each interpreter to jointly execute programs defined across LLVM's intermediate representation (LLVM IR) and Rust's mid-level intermediate representation (MIR). 
In Section~\ref{method:implementation}, we outline the architecture of our tool and describe how we resolved the differences in semantics between each interpreter. In Section~\ref{method:evaluation}, we describe our method for deduplicating test results to identify unique instances of undefined behavior and our approach to reporting bugs. 

\subsection{Sampling}
\label{method:sampling}
We evaluated our design on all compatible Rust libraries with test cases that called foreign functions. Rust libraries are referred to as ``crates,'' and they can be published at \href{https://crates.io/}{crates.io}. We used a snapshot of the crates.io database taken on September 20th, 2023. It contained 125,804 unique crates, of which 96\% (121,015) had at least one valid published version. We compiled each crate using version 1.74 (\href{https://github.com/rust-lang/rust/commit/37390d65636dd67e263753a3c04fbc60dcc4348e}{nightly-2023-09-25}) of the Rust toolchain, matching the version used by \toolname. This was necessary to ensure that we could use an unmodified version of Miri as a control to identify tests that would normally fail due to lack of support for foreign function calls. It would have been infeasible to maintain separate versions of \toolname for multiple toolchains without compromising the consistency of our results, since Miri depends on APIs that changed significantly throughout our evaluation. There is no standard method for using a version of Miri compiled with one toolchain to execute programs compiled with another toolchain, and there is no guarantee that Miri would behave as expected under these conditions.

Of all valid crates, 67\% (84,106) compiled without intervention. Of the crates that compiled, 36\% (44,661) had unit tests and 9\% (11,120) produced LLVM bitcode files, leaving 3\% (3,785) of crates with both unit tests and LLVM bitcode. Then, we used an unmodified version of Miri to execute all of the unit tests from this subset of crates to determine which tests called foreign functions. Of the 88,637 tests that we identified, 53\% (47,189) passed, 41\% (36,766)  failed, 4\% (3,869) timed out after five minutes, and 1\% (1,178) had been manually disabled. Tests can be disabled using the \code{#[ignore]} attribute or with conditional compilation directives.

Of the tests that failed in Miri, 63\% (23,116) failed due to foreign function calls. We executed this subset under both Stacked and Tree Borrows using an initial build of our tool to determine which tests called foreign functions. Of all potentially viable test cases, 39\% (9,130) called a foreign function we could execute. These tests originated from 25\% (957) of the crates with both test cases and bitcode. Our sample only includes crates that statically link to foreign code in their default configuration; it does not include crates that default to dynamic linking or statically link in non-default configurations. However, the 957 crates that we identified were more than enough to find meaningful answers to our research questions. We used this sample to conduct our final evaluation.

We also partially measured the distribution of crates that interact with foreign code. When we compiled each crate, we used an early lint pass to examine the abstract syntax tree and record the locations of all function declarations with a foreign ABI. We used data from this pass to identify 2,516 crates that declared a foreign function. This represents only 2.1\% of all valid crates. However, 12,564 (10.4\%) of all valid crates had a version that directly depended on at least one prior version of a crate that declared a foreign function binding. The maximum number of dependents for any crate that declared foreign functions was 6,411, with a mean of 11.0 dependents and a standard deviation of 171.8 dependents. 

\subsection{Implementation}
\label{method:implementation}
\begin{figure}
\caption{The architecture of \toolname.}
\label{figure:system}
\centering
\includegraphics[scale=1.5]{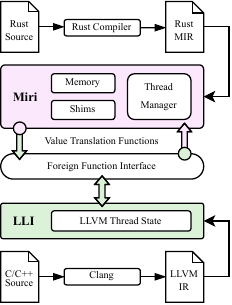}
\end{figure}

We illustrate the data flow between each of the core components of \toolname in Figure~\ref{figure:system}. As indicated at the bottom of the Figure, \toolname requires Clang~\cite{clang} to be set as the default C and C++ compiler and configured to emit LLVM bitcode during the build process for the library under test. Other than this requirement, using \toolname is identical to using an unmodified version of Miri.
 
When \toolname encounters a foreign function call in a Rust program, it looks for a corresponding definition in the LLVM module.
After finding a definition, the function's arguments are passed through a translation function, which lowers them into the representation used by LLI. There are two translation functions indicated by the arrows in the middle of Figure~\ref{figure:system}; one from Miri's representation to LLI's (left) and another from LLI's representation to Miri's (right). Miri passes the translated arguments to LLI, which creates a new thread to execute the function. Then, Miri sets the current current Rust thread to join on this new LLVM thread. After the LLVM thread terminates, its return value is passed back through the translation layer and given to the Rust thread, which continues to execute. A similar process is used when an LLVM thread calls a Rust function. Although Miri is single-threaded, it supports concurrency by non-deterministically stepping through multiple simulated ``threads'' of execution. LLI did not originally support any form of multithreading, but we modified it to be compatible with Miri's implementation. We support calling foreign functions from parallel Rust threads, but we do not support multithreading within LLVM. We implemented the Rust interface to LLI as an extension of Inkwell: a high-level Rust encapsulation for LLVM~\cite{inkwell}.

Nearly all of Miri's mechanisms for detecting undefined behavior are built into the core functions of the interpreter and do not require additional instrumentation of the source program. However, to detect aliasing violations, Miri configures the Rust compiler to insert ``retag'' instructions, which are used by its implementations of Stacked and Tree Borrows. We did not need to insert these instructions into the LLVM IR of foreign functions, since we treat LLVM pointers as equivalent to Rust's raw pointers.

We enabled LLI to detect undefined behavior by replacing several of its core operations with foreign function calls that use Miri to execute equivalent operations. As indicated by the labels next to ``Miri'' in Figure~\ref{figure:system}, all operations related to managing threads, accessing memory, and executing ``shim'' implementations of intrinsics and system calls are handled by Miri---for both interpreters. LLI manages its own local state for LLVM threads and implements the interpretation functions for each instruction, but Miri controls the process of taking a step for each thread. This architecture allowed us to extend Miri to detect errors in foreign function calls ``for free'' without having to reimplement its dynamic analyses within LLI. However, we had to address two additional difficulties to increase test coverage.

\paragraph{Translation}

Rust's MIR uses different calling conventions and represents values differently than LLVM IR, so we could not rely on a function's definition to have the same signature as its binding. 
However, foreign function bindings are also allowed to be incorrect, so we could not wholly trust type information to guide our conversion, either. Additionally, neither Rust nor LLVM has a formal model of its ABI that we could use as an oracle. When implementing our value translation functions, 
we settled on a conservative approach that supports the most common ABI differences while still being able to detect certain instances of incorrect bindings. We maintain the invariant that the size of a typed value on one side of the boundary must be equal to the size of the type expected by the other side. We provide a formal model of our value conversion functions in Section 2 of the Appendix. It demonstrates that conversion will fail when a difference in size is detected, which we report as undefined behavior.

Conversion is trivial for primitive types, since they have a canonical representation in Rust and LLVM. We also allow implicit casts between pointers and integers at foreign boundaries, which was necessary for 526 test cases from 70 crates. When aggregate types are passed by value, we require both the size and number of fields to match unless the aggregate is passed as an integer or by reference. However, we also frequently observed functions that expected individual aggregates to pass through multiple parameters, with one parameter for each field. This occurred only with homogeneous aggregates, where every field is identical and lacks padding. To accommodate this, we treat each field in a homogeneous aggregate as a separate argument if the entire aggregate cannot be passed as a single parameter. We apply this transformation only if the number of fixed parameters is greater than or equal to the number of remaining arguments plus the number of fields in the aggregate. This was necessary for 2\% (156) of our test cases from 15\% (140) of crates.

Products lack a canonical form between Rust and LLVM, so we do not support passing them as variadic arguments to LLVM or as arguments from LLVM to Miri's shims for system calls. Each of these types of functions lacks the type information that we need to guide our conversion. When an opaque pointer from LLVM is passed to one of Miri's shims, we assign it the type \code{*mut u8}. However, if it points to stack or static memory with a size equal to that of a primitive type, we assign it to point to that type. This may lead to false positive alignment errors when we cannot resolve an underlying type, and Rust expects a type other than \code{u8}.

\paragraph{Initialization}
Miri tracks which bits are initialized in each allocation and reports an error when uninitialized memory is read. 
However, LLVM will read uninitialized memory for the purpose of propagating \verb|undef| and \verb|poison| values, which represent the result of an indeterminate or invalid computation. 
If a single bit needs to be set within an uninitialized byte, instead of writing a zeroed byte with the individual bit set, LLVM will load the uninitialized byte, set only that bit, and then write it back~\cite{mciver22, juneyoung17}. 
These accesses are considered undefined behavior by Miri, but we treat them as false positives. 
We implemented two modes that allow us to identify true positive uninitialized reads of Rust-allocated memory in LLVM. By default, we allow \verb|load| instructions to read uninitialized bytes. However, \toolname can also be configured to treat uninitialized reads as errors while zero-initializing all LLVM-allocated stack and heap memory. Tests that read uninitialized memory must be run in each mode to ensure a complete evaluation. 
We cannot detect true positive uninitialized reads that occur outside of interactions with Rust, but these errors can be found with other tools~\cite{seward05, llvm} and are beyond the scope of our evaluation.

\subsection{Evaluation}
\label{method:evaluation}
We used \toolname to execute each of the 9,130 viable test cases that we identified in Section~\ref{method:sampling}. The version of \toolname at this point was based on the same version of the Rust's nightly toolchain (1.74.0) that we used in earlier stages of data collection. We set the global C and C++ compilers to version 16 of Clang and disabled optimizations. We collected data using Amazon EC2 on-demand instances provisioned through CloudBank~\cite{cloudbank}. We used \verb|c6a.2xlarge| instances during the initial stages of data collection, but we switched to \verb|c6a.xlarge| instances when using \toolname. It is single-threaded, so fewer resources were necessary. All commands were executed in a Docker container running Ubuntu 23.04.

\paragraph{Executing Tests} We executed each viable test natively and in \toolname under both Stacked Borrows and Tree Borrows. We configured Tree Borrows to treat values of the \code{Unique<T>} type as having the semantics of a mutable borrow. Without this configuration option, only the values of type \code{Unique<T>} within \code{Box<T>} are treated this way. We disabled isolation to allow executing non-deterministic operations and enabled symbolic alignment checking. By default, Miri will report an error if the address of a pointer is not a multiple of the requested alignment of the type of value being read or written through it. However, the base address of an allocation is not guaranteed to be a multiple of the requested alignment, so it is possible for a misaligned pointer to be ``aligned'' by chance. Symbolic checking avoids these false negatives by ensuring that the pointer's offset from the base address is a multiple of the alignment of the value being read and that the alignment of the allocation is greater than the alignment of the value. We ignored unaligned accesses in LLVM to prioritize detecting Rust-specific errors. 

On average, across each memory mode, 61\% of tests terminated due to an unsupported operation, 19\% passed, 10\% timed out, 1\% failed, and 9\% had a potential bug. A ``bug'' includes both undefined behavior and other undesired behaviors, such as memory leaks.
Of the 61\% of unsupported operations, 56\% were dynamically linked functions and inline assembly, which are out of scope. An additional 27\% were atomic instructions and floating point types that LLI does not implement, 12\% were due to types that are not supported by our value conversion layer (discussed in \ref{method:implementation}), 
and 5\% were shims that Miri does not fully implement. 

\paragraph{Deduplicating Errors} We reason about bugs in terms of test outcomes, where an outcome includes the results of a single test under both aliasing models. We deduplicated outcomes based on exit codes, stack traces, and error logs to avoid filing redundant bug reports. Prior to deduplication, we modified the error logs to remove unnecessary elements, such as references to specific memory addresses. We also included more or less detail in stack traces, depending on the location of an error. For foreign errors, we used the subset of the stack trace up to the Rust boundary to avoid deduplicating errors that appeared to be identical but were caused by mistakes at different callsites. When an error occurred in Rust, we only used the first line of the stack trace, since few of these errors were caused by LLVM. 

After deduplication, we had 394\xspace errors to investigate. Multiple factors lead to the number of bugs being significantly smaller than the number of deduplicated test outcomes. Our method of deduplication was conservative, we ignored errors from crates that had been unpublished since the start of our investigation, and we only reported Stacked Borrows violations that were also Tree Borrows violations. We also observed false positives related to alignment and accesses through addresses that we do not emulate (e.g., \code{stdout}). Additional manual investigation was necessary to fully diagnose certain aliasing bugs (e.g. \refbug{littlefs2:1}), since MiriLLI currently does not record the source locations of valid updates to the state of the borrow tracker when they occur in LLVM. Miri typically provides a trace of the history of each permission involved in a aliasing violation; only partial traces were available for certain bugs.

\paragraph{Reporting Errors} Not all errors are undefined behavior, and not all instances of undefined behavior are readily exploitable. However, we still attempted to follow ethical vulnerability disclosure practices by reporting bugs privately via email before creating public reports. When we found a bug, we examined its crate's \verb|Cargo.toml| and the GitHub profile of its repository's owner to find a way to contact its maintainers privately. If we were unsuccessful, we logged a public issue. We also logged issues if we had not received a response after at least one month or if the type of error did not appear to be exploitable. Our reports typically included a representative test case, the output from \toolname, and a minimal example where applicable. When a fix was trivial, we also filed a pull request. Section 1 of our Appendix includes metadata for each library and links to our contributions.

\section{Results}
\label{section:results}
We found 46 bugs in 37 crates. All appear to be new. As of December 13th, 2024, 28 of the bugs have been confirmed and fixed, and none have been identified as security vulnerabilities. Bugs occurred slightly more frequently in LLVM than in Rust; 23 occurred in LLVM, 17 occurred in Rust, and six were related to incorrect bindings. The majority (32) were found in crates with less than 100 average daily downloads in the six months prior to our snapshot of crates.io. However, we discovered three Tree Borrows violations in separate crates that each had more than 10,000 average daily downloads. We found several additional errors that occurred in Rust and did not meaningfully involve foreign code. These errors are excluded from the counts here, but they appear in our raw dataset.

We found 90 tests from 37 crates where Stacked Borrows violations occurred in foreign code. However, 66\% (59) of these tests passed or encountered an unsupported operation under Tree Borrows. This is mainly because 50\% (45) had invalid range errors similar to the minimal example shown earlier in Figure~\ref{figures:borrows:range}. For example, we observed that slices of an array would be cast into raw pointers, passed across the FFI, and then offset beyond the range that they initially borrowed. This is considered undefined behavior under Stacked Borrows, but it is valid under Tree Borrows. This provides our answer to \textbf{RQ2}; Tree Borrows accepts more real-world programs that call foreign functions than Stacked Borrows due to differences in handling pointer arithmetic.

\begin{table}
\scriptsize
\caption{Counts of each unique error grouped by category, the location of the fix, and the location of the error.}
\label{results:table:bugcounts}
\centering

\begin{tabular}{cccccc}
\multicolumn{1}{l}{}            & \multicolumn{1}{l}{}                         & \multicolumn{3}{c}{\textit{Category}}                                                                                  & \multicolumn{1}{l}{}     \\
\textbf{Fix}                    & \textbf{Error}                               & \multicolumn{1}{l}{\textbf{Allocation}} & \multicolumn{1}{l}{\textbf{Ownership}} & \multicolumn{1}{l}{\textbf{Typing}} & \textit{\textbf{Total:}} \\ \hline
\rowcolor[HTML]{FFFFFF} Binding & Binding                                      & -                                       & -                                      & 6                                   & 6                        \\
\rowcolor[HTML]{EFEFEF} Binding & LLVM                                         & -                                       & 3                                      & -                                   & 3                        \\
\rowcolor[HTML]{FFFFFF} LLVM    & LLVM                                         & -                                       & 3                                      & -                                   & 3                        \\
\rowcolor[HTML]{EFEFEF} Rust    & LLVM                                         & 1                                       & 16                                     & -                                   & 17                       \\
\rowcolor[HTML]{FFFFFF} Rust    & Rust                                         & 9                                       & 2                                      & 6                                   & 17                       \\ \hline
\multicolumn{1}{l}{}            & \multicolumn{1}{r}{\textit{\textbf{Total:}}} & 10                                      & 24                                     & 12                                  & 46                      
\end{tabular}

\end{table}

Now, we answer \textbf{RQ1} by identifying three categories of undefined or undesirable behavior that occur in Rust libraries that call foreign functions. We describe \textit{Ownership} errors in Section~\ref{bugs:ownership}, which include all violations of Tree Borrows and accesses out-of-bounds. We describe \textit{Typing} errors in Section~\ref{bugs:typing}, which include incorrect foreign function bindings and uninitialized values. We describe \textit{Allocation} errors in Section~\ref{bugs:allocation}, which include both memory leaks and cross-language deallocation.  Table~\ref{results:table:bugcounts} shows the number of bugs in each category, grouped by the location of the error and the location where a fix would need to be applied. We refer to each bug using a unique numerical ID corresponding to tables in Section 1 of the Appendix.

\subsection{Ownership}
\label{bugs:ownership}
We found 24 Ownership bugs from 20 crates. These include 17 Tree Borrows violations and seven accesses out-of-bounds.

\paragraph{Const-Correctness}
We found 10 insufficient permission errors caused by incorrectly casting immutable references to mutable raw pointers. In Bug~\refbug{klu-rs:1}, a chain of casts was used to convert \code{&self} into a mutable raw pointer. This pointer still had read-only \tbFrozen permission, so mutating through it across the FFI triggered an insufficient permission error. There were seven bugs with equivalent errors. The remaining three bugs occurred due to foreign function bindings that incorrectly declared pointers as \code{const} instead of \code{mut}, leading developers to pass pointers derived from immutable references instead of mutable ones.
\begin{figure}
\centering
\caption{A minimal example of Bug~\refbug{littlefs2:1}: an incorrect encapsulation of a self-referential pattern.}
\label{results:tb:selfreferential}
\scriptsize
\centering
\input{figures/self_referential.tex}
\end{figure}
    
\paragraph{Self-Reference}
In Rust, it is possible, but nontrivial, to implement self-referential patterns using safe encapsulations of unsafe operations~\cite{goregaokar22}. Bug~\refbug{littlefs2:1} demonstrates how a self-referential pattern can be implemented incorrectly. We found this bug in an encapsulation of a C library implementing a minimal file system for embedded applications. It used separate structs to represent the state of the file system and its configuration. The configuration held a mutable reference to the state, but both objects were contained in a parent struct representing the entire file system. Rust held a single mutable reference to the parent, which became invalid when the state was mutated through the configuration by a foreign function. 

Figure~\ref{results:tb:selfreferential} provides a minimal example of the bug and our fix. Lines we removed are highlighted in {\btHL[fill=hlcolorred]red} and marked with ``-'', while lines we added are highlighted in {\btHL[fill=hlcolorgreen]green} and marked with a ``+''. Tags within a circle indicate the state of the tree prior to the fix, while tags within a square represent the state of the tree after the fix. When a tag is on a line, it indicates that the pointer or reference on that line has a permission within the subtree corresponding to the tag. The tables at the bottom display the state of the tree before and after the write access on line~\ref{sr:mutate}.

The Rust encapsulation for \code{open} initializes the field \code{buffer} of the struct \code{Alloc} on line~\ref{sr:create} with a mutable raw pointer cast from a mutable reference to the field \code{cache}. The struct is mutably borrowed again on line~\ref{sr:again}. Each of these borrows, indicated by \circled{2} and \circled{3}, corresponds to an adjacent \tbReserved branch in the tree. The reference assigned to \code{b}, which has permission \circled{3}, is used to access the field \code{buffer}, which has permission \circled{2}. The pointer in this field is passed into the foreign function \code{ffi::open}.
Across the FFI, a write access occurs on line~\ref{sr:mutate} using permission \circled{2}. This is a child access relative to $\circled{1}$ and $\circled{2}$, so they transition to \tbActive. However, it is foreign access relative to $\circled{3}$, so this permission becomes \tbDisabled. This leads to an expired permission error on line~\ref{sr:ub} when $\circled{3}$ is used for a read access in Rust.

We fixed this bug by wrapping parts of the state in an \code{UnsafeCell}, which can be mutated through shared references. In Figure~\ref{results:tb:selfreferential}, we change the first field of \code{Alloc} to \code{UnsafeCell<i32>} and replace the mutable borrow assigned to \code{a.buffer} with \code{a.cache.get()}. This expression does not perform a retag, so both \code{a} and \code{cache} share permission \rectangled{1}. Since \code{a} contains an \code{UnsafeCell}, the mutable borrow assigned to \code{b} on line~\ref{sr:again} associates the tag $\rectangled{2}$ with a special $\m{Reserved}^*$ permission that can tolerate foreign writes. Consequently, this permission is no longer disabled by the write through $\rectangled{1}$ on line~\ref{sr:mutate}, so the read access on line~\ref{sr:ub} is valid.
\begin{figure}
\centering
\caption{A minimal example of the incorrect encapsulations of cyclic aliasing patterns that we observed in Bugs~\refbug{bzip2:1} and~\refbug{flate2:1}.}
\label{results:tb:shared_mutability}
\input{figures/shared_mutability.tex}
\end{figure}

\paragraph{Multiple Mutable Aliasing} Expired permission errors occurred when pointers derived from mutable references were copied into the foreign heap. We found this type of bug in five separate libraries. The most notable examples were Bug~\refbug{bzip2:1} from \verb|bzip2| and Bug~\refbug{flate2:1} from \verb|flate2|, which are both popular compression libraries. In particular, \verb|flate2| is actively maintained by the Rust project and averaged more than 130,000 downloads per day during our observation period.

Figure~\ref{results:tb:shared_mutability} shows a minimal example of the cyclic aliasing pattern used by each library. We use the same notation as Figure~\ref{results:tb:selfreferential}. The structs \code{Stream} and \code{State} carry pointers to each other. Both are encapsulated in Rust by a \code{Compression} object, which carries a heap-allocated instance of \code{Stream} within a \code{Box}. The Rust function \code{new} allocates memory for a \code{Stream} object and passes a mutable reference carrying the \tbReserved permission $\circled{1}$ to \code{ffi::init}. This foreign function copies the pointer it receives to the \code{Stream} object into a newly allocated \code{State} instance on line~\ref{mm:cyclic}, creating a cyclic structure. Later, when the function \code{compress} is called, it creates an adjacent \tbReserved permission $\circled{2}$. Mutating through this permission on line~\ref{mm:mutate:1} causes \circled{1} to become \tbDisabled, making it undefined behavior to read using \circled{1} in the body of \code{ffi:compress} on line~\ref{mm:ub}. 

To fix each error, we unwrapped the \code{Box} using \code{into_raw} and stored a raw pointer to the allocation within the \code{Compression} struct. We modified the API so that this raw pointer would be used for every access to the allocation, ensuring that the cyclic structure would remain valid. This corresponds to permission $\rectangled{1}$ in Figure~\ref{results:tb:shared_mutability}, which remains \tbActive through each access. We also had to modify \code{Compression}'s implementation of \code{Drop} to rewrap this pointer using \code{Box::into_raw} to avoid a memory leak. 

Bug~\refbug{littlefs2:2} also involved a mutable reference being copied into the foreign heap, but we needed a different approach to fix it. It occurred in the same file system library as Bug~\refbug{littlefs2:1}, which avoided creating heap allocations in Rust as a constraint of the embedded application context. Instead of using a \code{Box<T>} to allocate the equivalent of the \code{Stream} object in Figure~\ref{results:tb:shared_mutability}, the Rust encapsulation passed around a mutable reference to a stack allocation. Similar to Bugs~\refbug{bzip2:1} and Bug~\refbug{flate2:1}, once the reference was created, it was passed to a foreign initialization function that copied the reference into a foreign heap allocation. After the initialization function returned, the mutable reference was moved into a \code{RefCell<&mut T>}. This performed a reborrow, creating a new \tbReserved permission as a child of the \tbReserved permission held by the foreign heap. Both of these permissions could be used interchangeably for read accesses but not for write accesses. If the foreign parent permission was mutated first, it would invalidate the child permission. If the child permission was mutated first, then both permissions would become \tbActive, but the next mutation through the parent permission would cause the child permission to become \tbFrozen, making future write accesses through the child permission undefined behavior. 

To fix this error, we stored the mutable borrow as a \code{RefCell<*mut T>}, which prevented new child permissions from being created. However, the type \code{T} was a struct, and certain foreign functions required pointers to its members. Reborrowing is typically necessary to create a reference to a member of a struct, but this creates a new permission, which we needed to avoid. Instead, we used Rust's \code{addr_of_mut!} macro, which creates a pointer to a location without borrowing it.\footnote{Rust \href{https://blog.rust-lang.org/2024/10/17/Rust-1.82.0.html}{1.82.0} provided native syntax: \code{&raw mut}} Together, these two changes ensure that all memory accesses on each side of the boundary use the same permission. The authors of Tree Borrows~\cite{treeborrows} found an equivalent bug in test cases for the Rust toolchain and applied the same fix. 

\paragraph{Protection} 
We found one protected permission error equivalent to the example in Figure~\ref{figures:borrows:protection}. In Bug~\refbug{tectonic:engine:bibtex:1}, a Rust function attempted to use a foreign function to grow a heap-allocated array of values. The foreign function was passed a pointer derived from a reference-type argument to the Rust encapsulation. Similar to Figure~\ref{figures:borrows:protection}, when this pointer was used to reallocate the array, its protected permission became \tbDisabled, which is undefined behavior. We fixed this bug by changing the function to receive a raw pointer instead. Raw pointers do not receive protectors, so they can be used for deallocation.

\paragraph{PhantomData}
Rust's \code{PhantomData<T>} behaves like an instance of \code{T}, but it can be created without providing a concrete value. However, a value of type \code{PhantomData<UnsafeCell<T>>} does not behave like an \code{UnsafeCell<T>}; it cannot be mutated through immutable references~\cite{jung20unsafecell}. In Bug~\refbug{foreign-types:1}, the type \code{Opaque} was declared as an alias for \code{PhantomData<UnsafeCell<* mut T>>}. It was used to represent foreign types in Rust without having to declare their layout on both sides of the FFI boundary. Since \code{Opaque} did not contain a concrete instance of \code{UnsafeCell}, raw pointers derived from \code{&Opaque} received a read-only permission, leading to an insufficient permission error when they were used for a write access in foreign code. We proposed fixing this error by removing \code{PhantomData}, but other contributors are still discussing the correct fix. The relationship between \code{PhantomData} and \code{UnsafeCell} is not yet settled, so this pattern may be allowed in the future.

\paragraph{Access Out-of-Bounds}
We found seven access out-of-bounds errors in unique crates. Bugs~\refbug{dec:2} and~\refbug{dec-number-sys:3} were found in two encapsulations for \verb|libdecnumber|, which is part of the GCC toolchain~\cite{gcc}. This library often used the following pattern to iterate over arrays of integers:
\begin{rustfigurenoln}
for (; *curr==0 && curr+3 < end;) curr+=4;
\end{rustfigurenoln}
The first term in the conjunction will execute before the second term, but if the second term is false, then the first term is an access out-of-bounds or an uninitialized read. This was fixed by swapping the order of each term. Bug~\refbug{bad64:1} was found in a disassembler that incorrectly implemented several instructions, leading to an access out-of-bounds into adjacent static arrays. This error was fixed once the correct semantics were implemented. 

None of these errors meaningfully involve Rust. However, the remaining errors were caused by Rust encapsulations that did not adequately enforce the preconditions of foreign function calls. Bug~\refbug{crypto:pimitives:1} involved the following API, which redirected its arguments to foreign functions based on the value of the parameter \code{size}:
\begin{rustfigurenoln}
fn expand(ex_key: &mut [u8], key: &[u8], size: KeySize)
\end{rustfigurenoln}
For example, passing \code{KeySize::K128} would call a foreign function that assumed that \code{ex_key} and \code{key} would each be 128 bytes long. However, this invariant was not enforced by the encapsulation; it was possible to cause an access out-of-bounds error by passing a \code{KeySize} larger than the length of either key. We did not propose a fix for this bug, as it could require significant changes to the API.

Bug~\refbug{mseed:1} involved a foreign function that expected to receive a pointer to a string with six characters. It was allocated with \code{CString::new(Vec::with_capacity(6))}, exposed as a raw pointer using \code{into_raw}, and correctly deallocated after the call using \code{from_raw}. However, the implementation of \code{CString::new} appends a null-terminator to the vector and then converts it into a value of type \code{Box<[u8]>} using \code{into_boxed_slice}, which shortens the capacity of a vector to be equal to its length. The intended capacity was six, but \code{into_boxed_slice} shortened it to one, appending a null-terminator. The encapsulation did not validate that the length of the \code{CString} instance was correct, leading to an access out-of-bounds when the C codebase read beyond the first character. We fixed this error by initializing the \code{CString} with a constant of the correct length.

\subsection{Typing}
\label{bugs:typing}
We found 12 bugs from 12 crates that involved typing or initialization. Some were caused by incorrect use of \code{MaybeUninit<T>}, while others involved incorrect foreign function bindings.  

\paragraph{Incomplete Initialization} We found five bugs that involved partial initialization. In Bug~\refbug{dec:3}, an uninitialized instance of a struct \code{T} containing an array of values was created using \code{MaybeUninit<T>} and passed across the FFI to be initialized. The foreign call only initialized the first element of the array to zero. The remaining uninitialized elements would never be read by C, since every iteration stopped at the null terminator. Miri reported these initialization patterns as an error when \code{MaybeUninit<T>::assume_init} was called. We fixed this bug by zero-initializing the entire array in Rust.

Bug~\refbug{xxhrs:1} seemed to involve an equivalent pattern; \code{MaybeUninit<T>::assume_init} triggered an error for a struct that was initialized by a foreign function. However, the function properly zero-initialized the struct using \code{memset}. After examining the LLVM bytecode of the foreign function, we found that all 88 bytes of the struct were initialized on the LLVM stack, but only the first 80 bytes were copied back into Rust. These missing bytes corresponded to padding fields, which were never directly accessed by the foreign library. Optimizations had been disabled, but LLVM still removed this unnecessary write access. We fixed Bug~\refbug{xxhrs:1} by zero-initializing the padding fields before calling \code{assume_init}.

\paragraph{Incorrect Bindings}
We found six crates with one or more incorrect foreign function bindings. All of these incorrect bindings had been written manually. Three of these crates had bindings with missing return types. In Bug~\refbug{spritz:cipher:1}, a binding was declared in a unit test without its 32-bit integer return type. The C implementation used this to return a status code indicating if one of its integer parameters was within bounds. The remaining three bugs involved bindings with incorrect integer types, which led to incomplete initialization. In Bug~\refbug{special-fun:1}, a function was declared to return a 32-bit integer, but its C implementation returned a boolean value. In Bug~\refbug{secp256k1:1}, the last parameter of a function was declared as a 32-bit integer, but the C implementation expected \code{size_t}. The width of \code{size_t} is architecture-dependent, so this binding was only correct for 32-bit architectures. To fix this bug, multiple prior releases needed to be patched to use Rust's \code{usize} type, which has similar semantics. 

\subsection{Allocation}
\label{bugs:allocation}
We found 10 issues related to allocation in 10 distinct libraries. These include nine memory leaks and one new cross-language deallocation bug.

\paragraph{Memory Leaks} Rust's \code{Box<T>} and \code{CString} use similar APIs to encapsulate heap allocations. For each type, the function \code{into_raw} consumes an instance and produces a pointer to its interior heap allocation. To avoid a memory leak, this pointer must be reconstituted in its wrapper type using the function \code{from_raw}. We found four leaks caused by calling \code{into_raw} on either \code{Box<T>} or \code{CString} without later calling \code{from_raw}. Each type was used to allocate memory for a foreign function call, and each bug was fixed by adding a call to \code{from_raw} after the function returned. The remaining five leaks were of memory that had been allocated by C. Most were caused by neglecting to call a destructor that had been exposed by the C API. However, in Bug~\refbug{libcmark-sys:1}, the C API did not expose a destructor, since it had been designed based on the assumption that users would be able to call \code{free}. We could not use Rust's allocator API to fix this, since that could lead to invalid cross-language deallocation. 

\paragraph{Cross-Language Deallocation} It is potentially undefined behavior to free a pointer allocated by C in Rust, or vice versa, since each language's binary may use a different allocator. We found one new example of cross-language deallocation. In Bug~\refbug{quickjs:regex:3}, a pointer to a heap-allocated string was returned to Rust by a foreign function call and stored as a \code{Cow<&'static [u8]>}, which will lazily clone its data when mutated. This wrapper type is an enumeration with two variants; \code{Cow::Borrowed} receives a reference to a value, while \code{Cow::Owned} takes ownership of a value, deallocating it when it goes out of scope. When this \code{Cow} was dropped, it deallocated the C heap memory using Rust's allocator. There was no immediate fix for this bug, since the C API did not expose a destructor.

\section{Discussion}
\label{section:discussion}
Our findings indicate that it is easy to inadvertently introduce undefined behavior in Rust libraries that call foreign functions. The errors we found can be prevented with careful auditing, but it seems likely that many will persist until the Rust community develops a production-ready method for finding aliasing violations in multi-language applications. We provide the following recommendations:

\paragraph{For Rust Developers}
Awareness is key to avoiding improper use of unsafe features.
Developers who depend on foreign code should validate their tests with language-agnostic bug-finding tools, such as LLVM's sanitizers or Valgrind, which can detect memory leaks (e.g~\refbug{libcmark-sys:1}) and accesses out-of-bounds (e.g.~\refbug{bad64:1}). 
Developers who maintain foreign function bindings should consider that the infrequent cost of generating and committing bindings may be preferable to writing them by hand without assistance, which can lead to errors. When heap objects are accessed on each side of the foreign boundary, it is helpful to be aware of where each object is created, how many references it has, where these references are stored, and what capabilities they require. Each of these attributes can influence the correct design for an encapsulation. For example, Bug~\refbug{littlefs2:2} was fixable by casting a reference into a pointer, while Bugs~\refbug{flate2:1} and~\refbug{bzip2:1} required unwrapping a \code{Box}.

\paragraph{For The Rust Project} 
The Rust community is in dire need of a production-ready solution for finding aliasing bugs in multi-language applications. 
Although our approach was capable of finding real-world bugs, its scalability is limited due to the requirement of implementing shims for system calls and the lack of a formal specification for the ABIs implemented by Rust, Clang, and LLVM.
Instrumenting a shared, intermediate format is likely to be the most effective approach. The Krabcake project~\cite{krabcake} (announced concurrently with \toolname) is developing an extension to Valgrind~\cite{seward05} that will provide support for detecting aliasing violations, but it is not yet capable of replicating our results. The Rust community should invest resources into completing a prototype implementation to evaluate in real-world applications. However, Valgrind can also incur significant runtime overhead. Our future work will involve creating an LLVM sanitizer that can detect aliasing violations, providing the performance necessary to find these errors at scale with fuzzing tools. We also expect that sophisticated static analysis tools would help developers avoid undefined behavior at foreign boundaries by issuing warnings for functions that copy pointers or mutate through pointers derived from immutable references. 

\section{Related Work}
\label{section:related}
Foreign function calls are a common use case for unsafe code, but they have been understudied in prior work.

\paragraph{Surveys of the Rust Ecosystem} 
Studies that examine how unsafe code is used in Rust libraries have consistently found that foreign function calls are a significant use case for unsafe code~\cite{evans20,astrauskas20}.
In particular, Evans et al.~\cite{evans20} surveyed all published crates in September of 2018 and found that 22.5\% of all unsafe function calls were to foreign functions. 
Rust developers also view foreign function calls as a central use case for unsafe code. 
Fulton et al.~\cite{fulton21} interviewed 16 Rust developers and surveyed 178 developers, finding that nearly half of their participants used foreign function calls.
Both Höltervennhoff et al.~\cite{holtervennhoff23} and McCormack et al.~\cite{mccormack24} focused specifically on Rust developers who use unsafe code, and each found that the majority of participants used Rust's FFI.

\paragraph{Types of Bugs \& Undefined Behavior} 
Prior studies have examined open source contributions to categorize errors caused by unsafe code.
Most considered foreign function calls to be out of scope, and none described aliasing violations in terms of Stacked or Tree Borrows.
Both Qin et al.~\cite{qin20} and Xu et al.~\cite{xu22} examined bug and vulnerability reports from open source projects and found examples of allocation and typing errors that are similar to the ones that we discovered.
Xu et al.~\cite{xu22} did encounter errors related to foreign function calls, but they were primarily ``straightforward'' issues related to application-specific invariants, layout, and alignment.
Cui et al.~\cite{cui23} evaluated a taxonomy of 19 safety properties required by unsafe functions, but they made it an explicit design goal to exclude foreign functions.

\paragraph{Bug-finding Tools \& Verifiers} 
Prior approaches to static analysis~\cite{mirchecker, qin20, safedrop, rudra} have found a wide variety of bugs caused by improper use of unsafe code, but few approaches scale beyond Rust.
Li et al.~\cite{ffichecker} applied dataflow analysis on LLVM IR and detected several memory leaks and cross-language deallocation errors in Rust libraries. We replicated several of the bugs they encountered. 
Hu et al.~\cite{crust} modified several existing Rust analysis tools, including Miri, to analyze multi-language programs defined in a custom intermediate representation.
However, they did not consider aliasing violations.
Lei et al.~\cite{acorn} took a similar approach using WebAssembly as the target language, but their evaluation did not use Miri or consider aliasing issues, either. Recent deductive verifiers based on the RustBelt~\cite{rustbelt_jung18} model have used compositional verification~\cite{ayoun2024hybrid} and automated proof search~\cite{gaher24} to verify unsafe code, but none of these approaches reason about Stacked or Tree Borrows or have explicit support for foreign function calls. 

\section{Threats to Validity}
\label{section:threats}
\paragraph*{Construct Validity} We disabled compile-time optimizations in both Rust and LLVM to avoid missing bugs. 
However, we did not evaluate whether optimization would have had any effect on our results, and we did not determine whether any of the bugs that we found had an effect on native execution. 

\paragraph*{Internal Validity} We did not use the same hardware when investigating the source of each bug as we did for large-scale data collection. Our value translation layer cannot detect when a binding uses an incorrect type that happens to have the correct layout, which may have caused us to miss bugs. Errors in Miri or other components of Rust's nightly toolchain may have affected the correctness of our results. Since we manually investigated each bug, we have no reason to believe that these errors and their fixes are specific to any particular version of Rust's toolchain, with the exception of Tree Borrows violations. The model was still evolving during our evaluation. 

Several of our bug reports are still pending responses, so we cannot be entirely certain that all of our results are novel or true positives. However, we did exclude bugs from our evaluation that had already been reported. We did not independently confirm that these bugs were absent from test cases that did not call foreign functions, but we have no reason to believe that any would have been detected by an unmodified version of Miri at the time of our evaluation.

\paragraph*{External Validity} Crates that could not compile with our nightly toolchain were excluded from our evaluation, even if they would have otherwise been compatible with MiriLLI. Limited support for certain features, such as dynamically linked libraries, also prevented us from achieving high test coverage; 61\% of tests terminated due to an unsupported operation and 10\% timed out. Additionally, the majority of libraries we examined linked against C, and not C++. Crates that failed to compile using our nightly toolchain were excluded from the evaluation, as well as crates that overrode the system's default C and C++ compiler. However, these limitations did not prevent us from answering our research questions. Our goal was not to create a production-ready tool, but to understand the types of errors that can occur in Rust libraries that call foreign functions. The bugs that occur in general may differ in unexpected ways from those found with our setup.

\section{Conclusion}
\label{section:conclusion}
We conducted a large-scale evaluation to investigate the types of undefined behavior that occur in Rust libraries that call foreign functions. We created \toolname---a dynamic analysis tool that combines two existing interpreters---and we used it to find 46 bugs in 37 libraries. Though many of these bugs were aliasing violations, Rust's Tree Borrows aliasing model was more permissive in this context than the earlier Stacked Borrows model. To ensure that these errors are easy to detect, the Rust project should invest in new dynamic analysis tools that can accommodate multi-language applications.

\section*{Acknowledgment}
\label{section:acknowledgment}
We thank Tomas Dougan for contributing to the implementation of \toolname. This material is based on work supported by the Department of Defense and the National Science Foundation under Grant Nos. CCF-1901033, DGE1745016, and DGE2140739. Results presented in this paper were obtained using CloudBank~\cite{cloudbank}, which is supported by the National Science Foundation under Grant No. CNS-1925001. Any opinions, findings, conclusions, or recommendations expressed in this material are those of the authors and do not necessarily reflect the views of the Department of Defense or the National Science Foundation.
\bibliographystyle{IEEEtran}
\bibliography{refs}

\end{document}